\documentclass[a4paper]{jpconf}
\usepackage{graphicx}

\def\apj{ApJ}  
\def\aap{A\&A} 

\def\mnras{MNRAS} 
\begin{document}
\title{F-GAMMA: On the phenomenological classification of continuum radio spectra
  variability patterns of {\it Fermi} blazars}

\author{E~Angelakis$^{1}$, L~Fuhrmann$^{1}$, I~Nestoras$^{1}$, C~M~Fromm$^{1}$,
  M~Perucho$^{2}$, R~Schmidt$^{1}$, J~A~Zensus$^{1}$, N~Marchili$^{1}$,
  T~P~Krichbaum$^{1}$, H~Ungerechts$^{3}$, A~Sievers$^{3}$, D~Riquelme$^{3}$ and V~Pavlidou$^{1}$}

\address{$^{1}$ Max-Planck-Institut f\"ur Radioastronomie, Auf dem H\"ugel 69, DE-53121,  Bonn, Germany}
\address{$^{2}$ Department d'Astronomia i Astrof\'{i}sica, Universitat de Val\`{e}ncia, C/Dr. Moliner 50, 46100 Burjassot, Val\`{e}ncia, Spain}
\address{$^{3}$ Instituto de Radio Astronomía Milim\'{e}trica, Avenida Divina Pastora 7, Local 20, E 18012, Granada, Spain}

\ead{eangelakis@mpifr.de}

\begin{abstract}
  The {\em F-GAMMA} program is a coordinated effort to investigate the physics of Active
  Galactic Nuclei (AGNs) via multi-frequency monitoring of {\em Fermi} blazars. In the
  current study we show and discuss the evolution of broad-band radio spectra, which are
  measured at ten frequencies between 2.64 and 142\,GHz using the Effelsberg 100-m and the
  IRAM 30-m telescopes. It is shown that any of the 78 sources studied can be classified
  in terms of their variability characteristics in merely 5 types of variability. It is
  argued that these can be attributed to only two classes of variability mechanisms. The
  first four types are dominated by spectral evolution and can be described by a simple
  two-component system composed of: (a) a steep quiescent spectral component from a large
  scale jet and (b) a time evolving flare component following the ``Shock-in-Jet''
  evolutionary path. The fifth type is characterised by an achromatic change of the broad
  band spectrum, which could be attributed to a different mechanism, likely involving
  differential Doppler boosting caused by geometrical effects. Here we present the
  classification, the assumed physical scenario and the results of calculations that have
  been performed for the spectral evolution of flares.
\end{abstract}

\section{Introduction}
Among the most evident characteristics of blazars is their intense variability at all
wavelengths. Studies of the variability characteristics, preferably with simultaneous
data, shed light on the physics driving the energy production and dissipation in these
systems (e.g. \cite{boettcher2010,boettcher2010HEAD}). The {\em F-GAMMA} program
\cite{fuhrmann2007AIPC,angelakis2008MmSAI..79.1042A,2010arXiv1006.5610A} explores this
possibility by coordinated monthly monitoring of {\em Fermi} $\gamma$-ray monitored
blazars. {\em F-GAMMA} is covering mostly the radio cm to sub-mm bands primarily with the
Effelsberg 100-m, the IRAM 30-m and the APEX 12-m telescopes (although optical telescopes
are participating as well, Fuhrmann et al. in prep.) for $\sim60$ prominent blazars.

The physical cause for the variability itself has been long debated. The ``Shock-in-Jet''
model suggested by \cite{Marscher1985ApJ}, is the most accepted one and attributes the
variability to shocks propagating down the jet. The basic assumption is that variations at
the jet base (e.g. changes in the injection rate, the magnetic field, bulk Lorentz factor
etc.) cause the formation of shocks, which consequently go through first {\em Compton},
then {\em synchrotron} and finally {\em adiabatic} energy losses. This is the main model
prediction which is adopted in the following.

An alternative model, the ``Internal Shock Model'' proposed by \cite{Spada2001MNRAS},
suggests that energy is channeled into the jet in an intermittent way. ``Plasma shells'',
may have different bulk Lorentz factors and masses, so that faster shells can catch up
with slower ones, collide and relativistic shocks can occur. The shock accelerates
electrons to relativistic energies so that they can emit synchrotron and inverse Compton
radiation.

Other models explain the variability geometrically. For example,
\cite{1999AnA...347...30V} suggest that orbital motion and jet precession, e.g. caused by
a binary black hole system, can produce helical jet morphologies which can bent the jet,
changing its orientation and shape. This may then cause brightness and spectral
variability through changes in the Doppler factor.

\section{Observations}
The work discussed here is based on observations conducted quasi-simultaneously with the
Effelsberg 100-m and the IRAM 30-m telescope within the {\em F-GAMMA} program which is
described in \cite{fuhrmann2007AIPC,angelakis2008MmSAI..79.1042A,2010arXiv1006.5610A}.
The 100-m telescope has been observing between 2.64 and 43.05\,GHz at 8 frequencies over a
baseline of 4.5 years (January 2007 -- June 2011). The 30-m telescope has been measuring
at 86 and 142\,GHz (June 2007 -- June 2011). The millimetre observations for {\em F-GAMMA}
are closely coordinated with the more general flux monitoring conducted by IRAM, and data
from both programs are included in this paper. The Effelsberg 100-m telescope is equipped
with circularly polarised feeds, while the IRAM 30-m with linearly polarised ones. Details
are given elsewhere (Fuhrmann et al. in prep., Angelakis et al. in prep., Nestoras et
al. in prep.). Measurements at 4.85\,GHz, 10.45\,GHz, 32.0\,GHz, 86.24\,GHz and 142\,GHz
are done differentially either by using multi-feed systems, or, at IRAM 30-m, by wobbler
switching.

On average, the time needed for observing an entire spectrum of any given source at
Effelsberg alone is of the order of 35 minutes, while at IRAM, roughly 2 minutes. The
combined spectra (Effelsberg and IRAM) are observed quasi-simultaneously within
approximately one week. That is, neither the single-facility nor the combined spectra are
likely to be affected by source variability. In the current study only data collected
until June 2011, have been used.

\section{Data reduction}
The antenna temperature measured for a certain source is subjected to a series of
corrections to compute the source flux density outside the terrestrial
atmosphere. These operations, are: (a) {\em Pointing correction}, meant to correct for
small pointing residuals. (b) {\em Elevation dependent gain correction}, correcting for
the loss of antenna sensitivity caused by gravitational deformation of the reflector. (c)
{\em Atmospheric opacity correction}, which corrects for the attenuation due to the
terrestrial atmospheric absorption. (d) {\em Absolute calibration (sensitivity
  correction)}, which does the translation of the previously corrected antenna temperature
to SI units. All measurements are subsequently exposed to quality checks. The overall
uncertainties are typically of the order of $0.5 - 5$\,\% for Effelsberg and of the order
of $\le 10$\,\% for IRAM. More details can be found in \cite{angelakis2009AnA} as well as
in Angelakis et al. (in prep.) and in Nestoras et al. (in prep.).

\section{Phenomenological classification of the spectrum variability pattern}
\label{subsec:classification}
The following discussion refers to the phenomenological classification of the variability
patterns shown by the broad-band spectra of sources in the {\em F-GAMMA} sample relative
to the observing band-pass. The term ``variability pattern'' is assigned to the shape
traced by the source broad-band spectrum -- as that is observed monthly -- as a function
of time. In total 78 sources have been examined and their spectra span, in total,
over 4.5 years of observations.

\begin{figure}[h]
\begin{minipage}{12pc}
\includegraphics[width=0.7\textwidth,angle=-90]{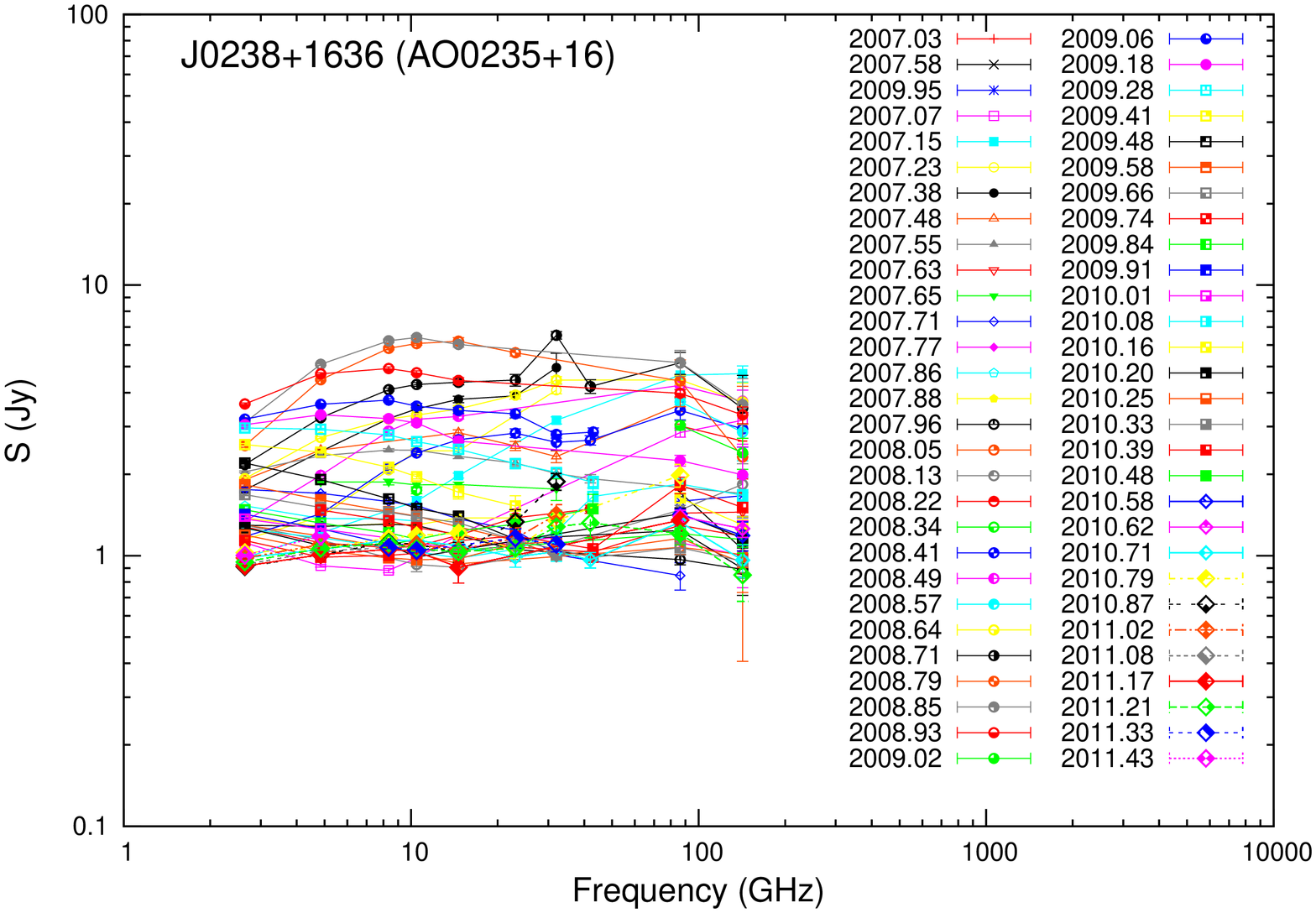}
\caption{\label{fig:t1}Prototype source for variability type 1.}
\end{minipage}\hspace{1pc}\vspace{1pc}%
\begin{minipage}{12pc}
\includegraphics[width=0.7\textwidth,angle=-90]{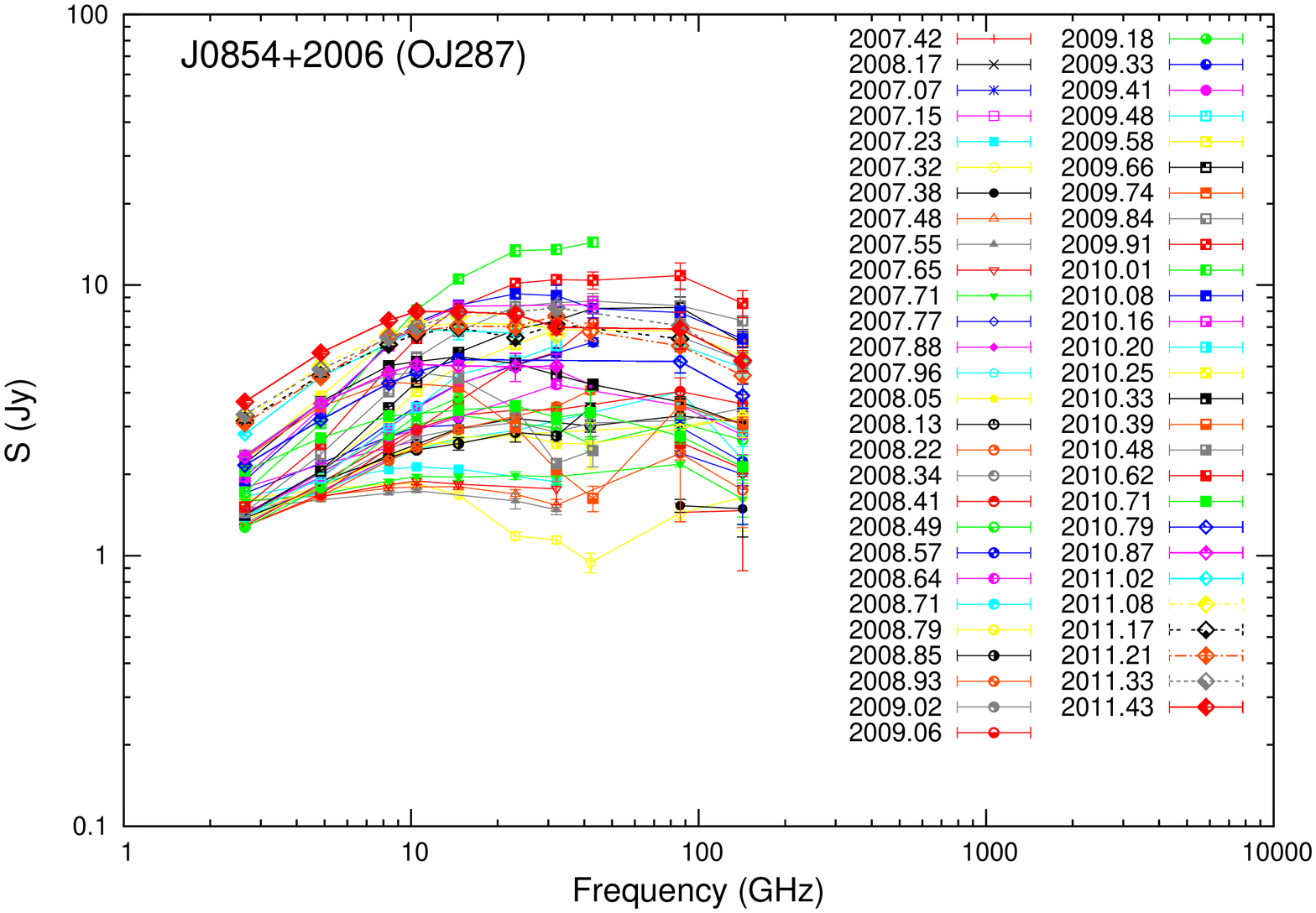}
\caption{\label{fig:t1b}Prototype source for variability type 1b.}
\end{minipage}\hspace{1pc}%
\begin{minipage}{12pc}
\includegraphics[width=0.7\textwidth,angle=-90]{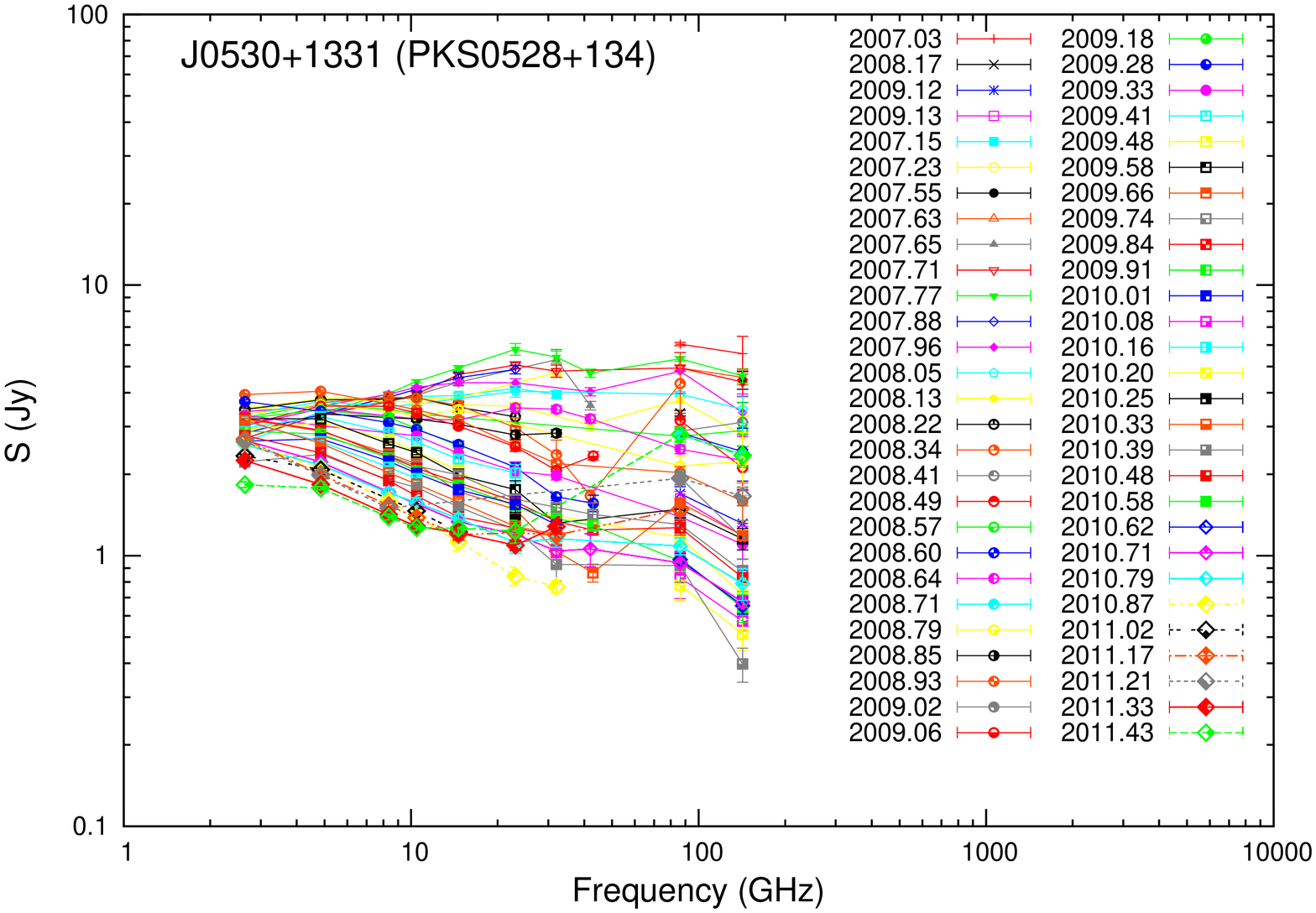}
\caption{\label{fig:t2}Prototype source for variability type 2.}
\end{minipage}\hspace{1pc}%
\begin{minipage}{12pc}
\includegraphics[width=0.7\textwidth,angle=-90]{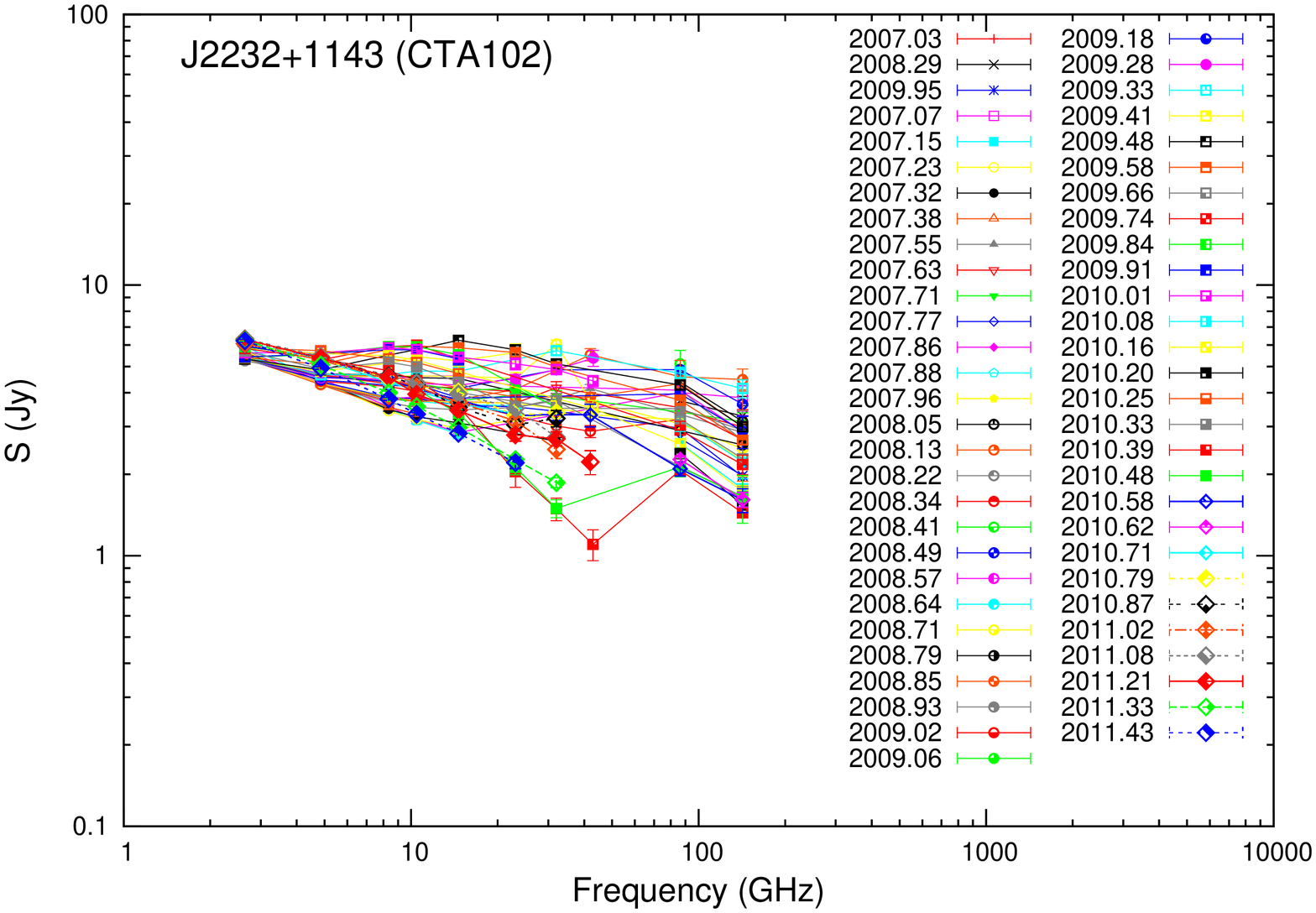}
\caption{\label{fig:t3}Prototype source for variability type 3.}
\end{minipage}\hspace{1pc}\vspace{1pc}%
\begin{minipage}{12pc}
\includegraphics[width=0.7\textwidth,angle=-90]{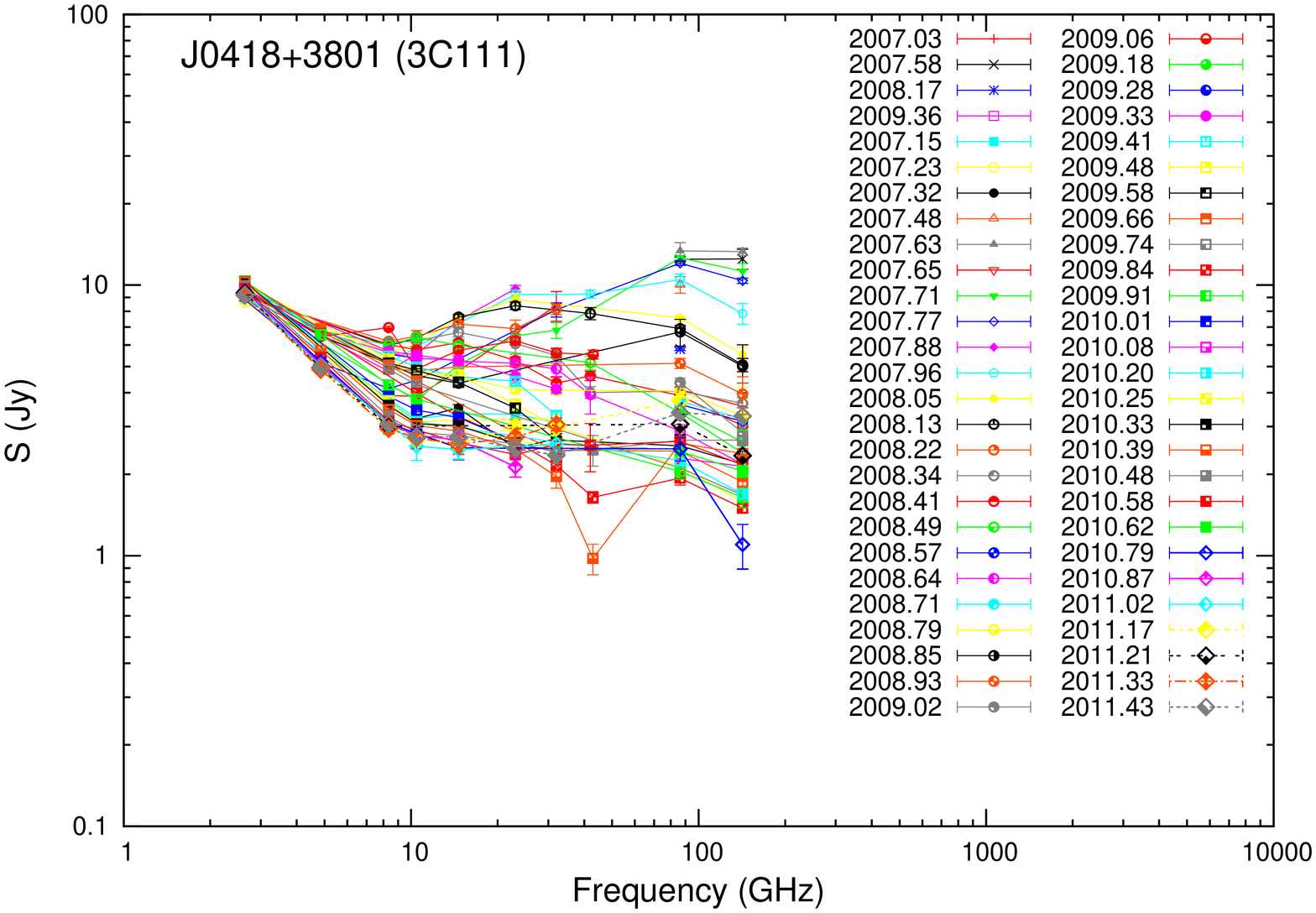}
\caption{\label{fig:t3b}Prototype source for variability type 3b.}
\end{minipage}\hspace{1pc}%
\begin{minipage}{12pc}
\includegraphics[width=0.7\textwidth,angle=-90]{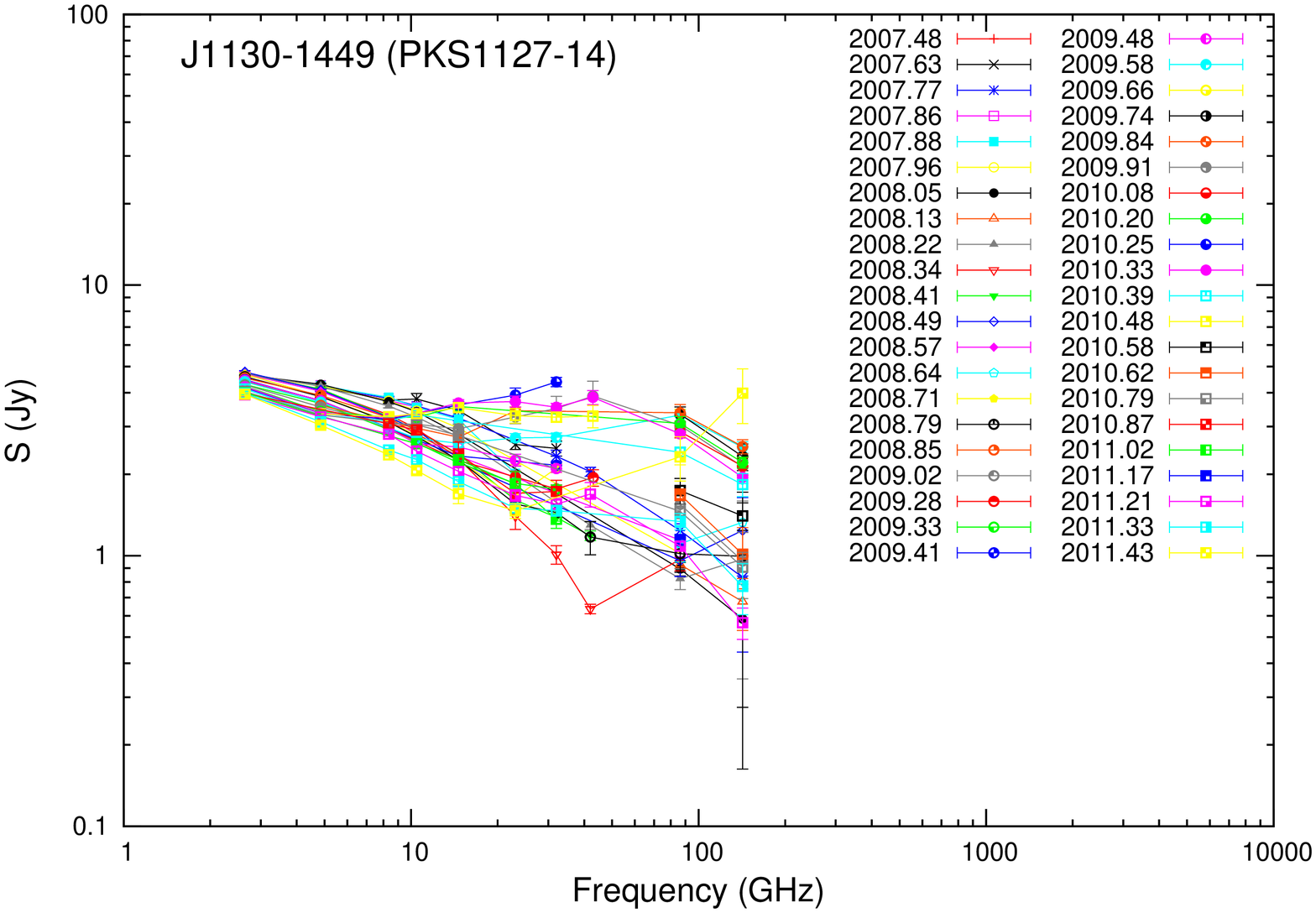}
\caption{\label{fig:t4}Prototype source for variability type 4.}
\end{minipage}\hspace{1pc}%
\begin{minipage}{12pc}
\includegraphics[width=0.7\textwidth,angle=-90]{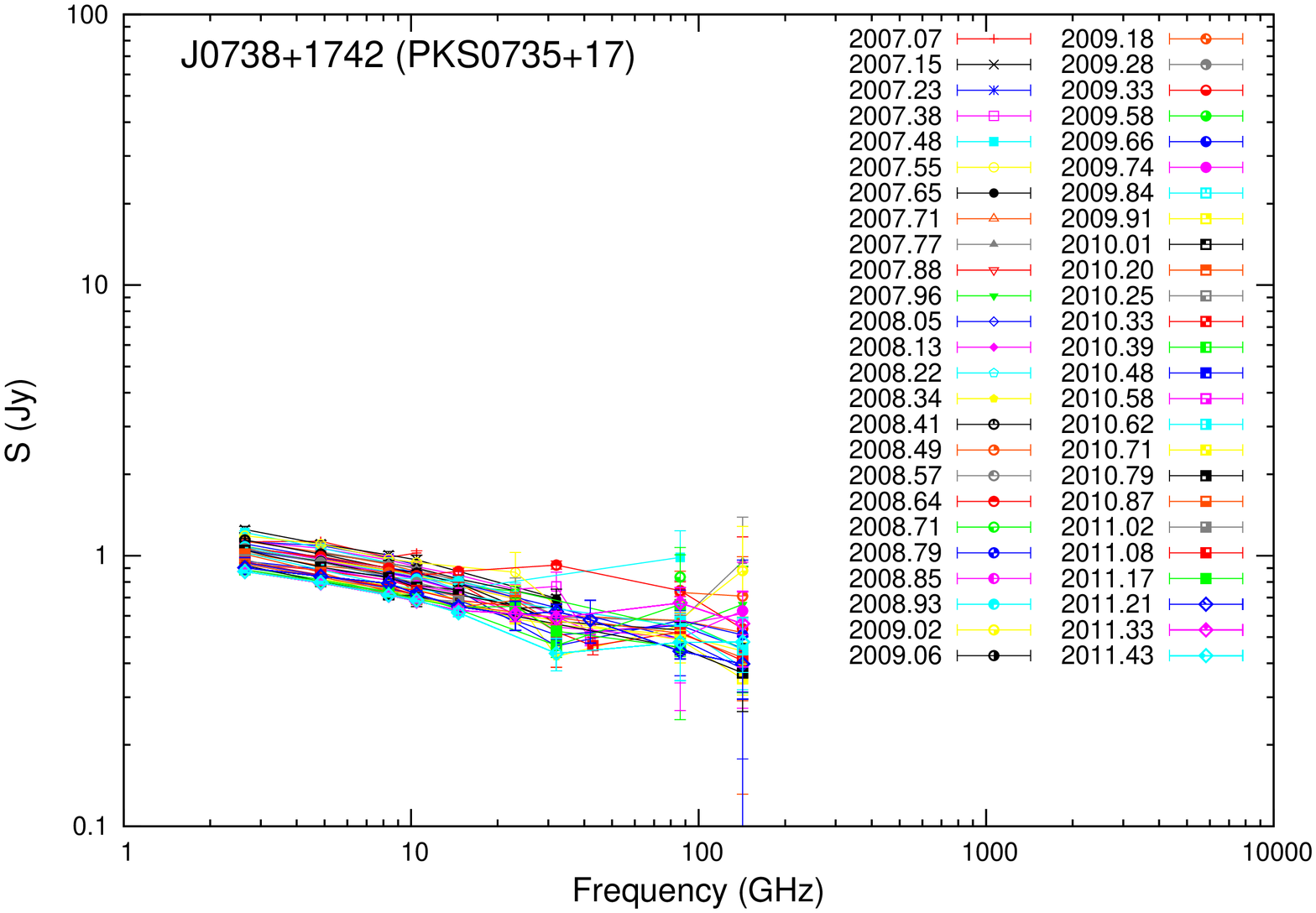}
\caption{\label{fig:t4b}Prototype source for variability type 4b.}
\end{minipage}\hspace{1pc}\vspace{1pc}%
\begin{minipage}{12pc}
\includegraphics[width=0.7\textwidth,angle=-90]{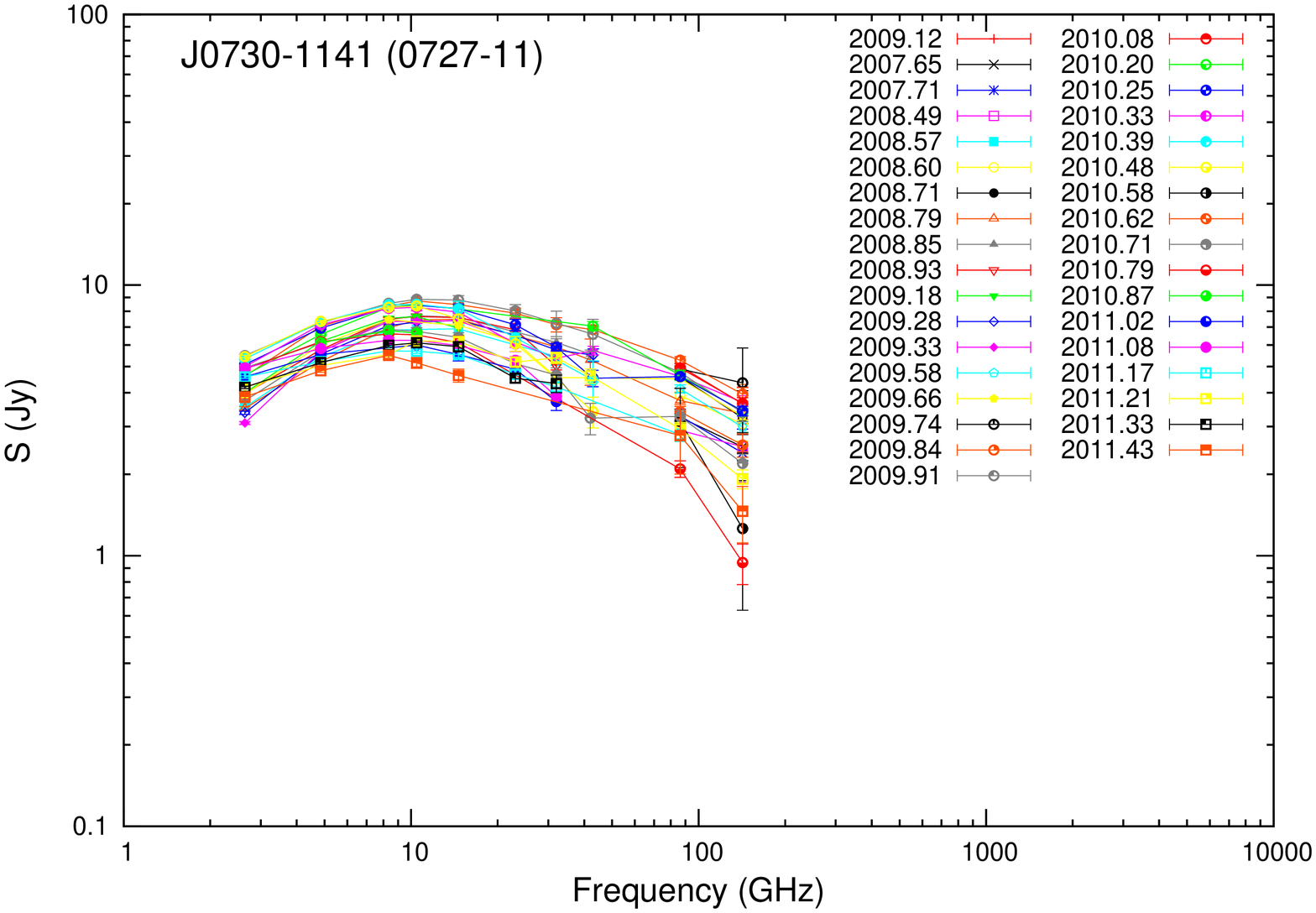}
\caption{\label{fig:t5}Prototype source for variability type 5.}
\end{minipage}\hspace{1pc}%
\begin{minipage}{12pc}
\includegraphics[width=0.7\textwidth,angle=-90]{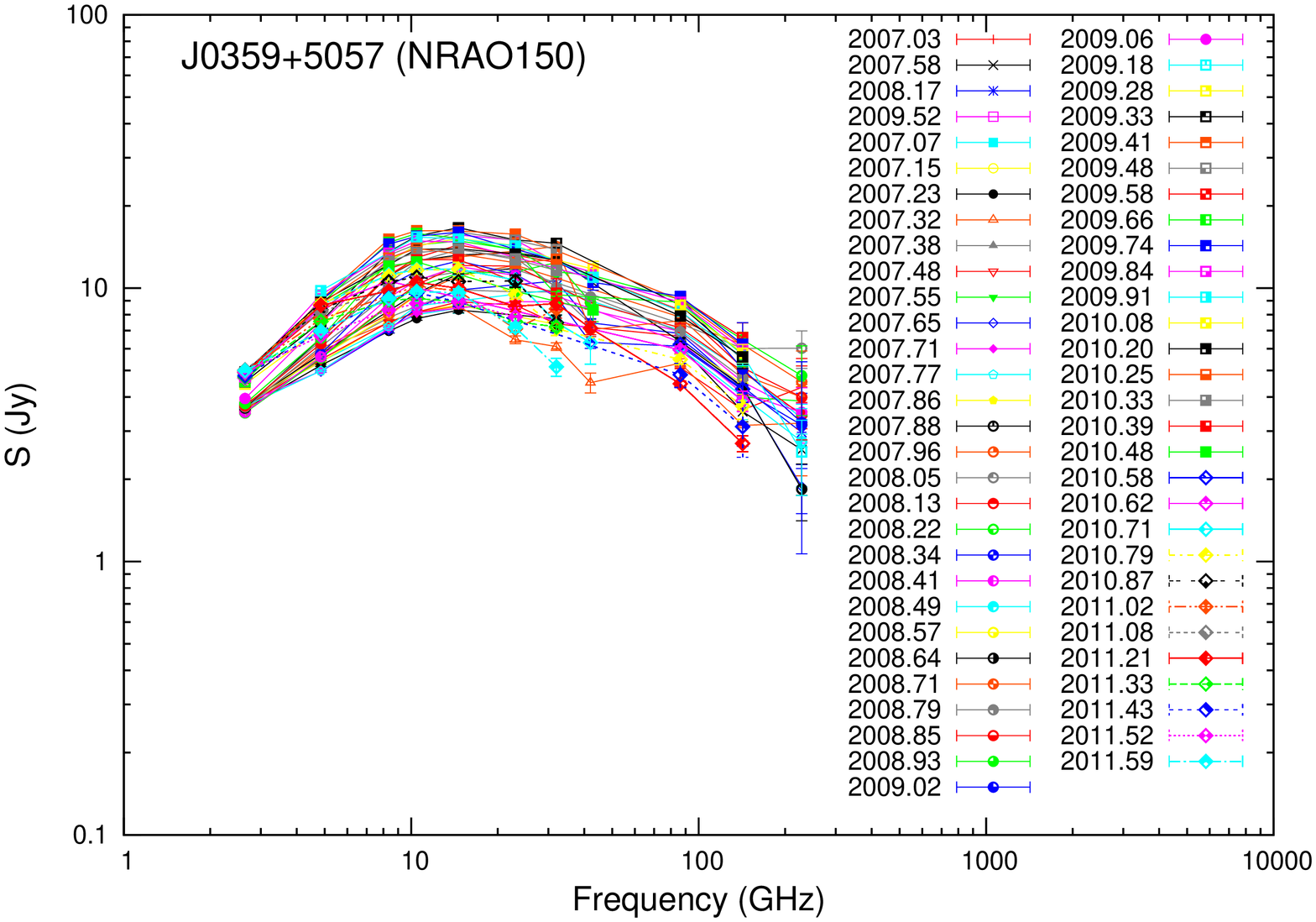}
\caption{\label{fig:t5b}Prototype source for variability type 5b.}
\end{minipage} 
\end{figure}

A visual inspection of the examined sources reveals a wealth of spectral features as well
as of variability patterns that different sources exhibit. For instance, the variability
amplitude, the frequency of occurrences and the pace at which the variability events
evolve, seem to be different for different sources. Interestingly, despite the
apparent complexity it appears that the 78 sources that have been studied here can
be classified in only five phenomenological classes on the basis of its
variability pattern, which are numbered from 1 to 5 (more details will be
given by Angelakis et al. in prep.). Four of them show also sub-types which however do not
deserve a separate type and are named after the main type followed by the letter
``b''. The prototype sources are shown in figures~\ref{fig:t1}--\ref{fig:t5b}. The main
phenomenological characteristics of these types, are:
\begin{description}
\item[Type 1]: This variability pattern is clearly dominated by spectral evolution. That
  is, at an instant of time the spectrum appears convex and its peak
  ($S_\mathrm{m},\nu_\mathrm{m}$) is drifting within the observing band-pass from high
  towards lower frequencies, covering a significant area in the $S-\nu$ space.  Its shape
  is smoothly changing towards an ultimate flat or mildly steep power law which is then
  followed by new events. For this type no evidence for the presence of an underlying
  quiescent steep spectrum is seen. The lowest frequencies in the bandpass are remarkably
  variable indicating that the activity ceases at frequencies much lower than the lowest
  in our band-pass. The prototype source is shown in figure.~\ref{fig:t1}.
\item[Type 1b]: As a sub-class of the previous one, type 1b shares the same
characteristics with type 1 except that the lowest frequency does not show as intense a
variability. The activity ceases around this part of the band-pass. A source
representative of this type is shown in figure~\ref{fig:t1b}.
\item[Type 2]: This type is also dominated by spectral evolution. The basic
characteristic of this case is the fact that the flux density at the lowest frequency
during the steepest spectrum phase is higher than that during the inverted spectrum
phase. Moreover, the maximum flux density reached by the flaring events is significantly
above that at the lowest frequency. This implies that the observed steep spectrum is not a
quiescent spectrum but rather the relic of an older, yet recent, outburst. The
prototype of this type is shown in figure~\ref{fig:t2}.
\item[Type 3]: Type 3, shown in figure~\ref{fig:t3}, is dominated by the spectral
  evolution as well. The identifying characteristics, are: (a) the fact that the lowest
  frequency practically does not vary and, (b) the maximum flux density level reached by
  outbursts is comparable to that at the lowest band-pass frequency. This phenomenology
  leaves hints that the events cease very close to the lowest frequency of the band-pass
  and hence a quiescent spectrum is becoming barely evident.
\item[Type 3b]: Type 3b, shown in figure~\ref{fig:t3b}, is very similar to type 3. Here
however the quiescent spectrum is seen clearly at least at the 2 lowest frequencies.
\item[Type 4]: Sources of this type spend most of the time as steep spectrum ones which
are sometimes showing an outburst of relatively low power propagating towards low
frequencies. A representative case is shown in figure~\ref{fig:t4}.
\item[ Type 4b]: This type includes persistently steep spectrum cases as it is shown in
figure~\ref{fig:t4b}.
\end{description}

In all previous classes, type 1 to 4b, the occurring variability is clearly dominated by spectral
evolution. However, there exists a class of sources that show a fundamentally different
behaviour. The variability happens self-similarly without signs of spectral evolution. Those
are grouped in a separate type with two sub-types:
\begin{description}
\item[Type 5]: In this case the spectrum is convex and follows an ``achromatic''
evolution. That is, it shifts its position in the $S-\nu$ space preserving its shape. This
is shown clearly in figure~\ref{fig:t5}.
\item[Type 5b]: This type shows, in principle, characteristics similar to the previous
one but there occurs a mild yet noticeable shift of the peak ($S_\mathrm{m},\nu_\mathrm{m}$)
towards lower frequencies as the peak flux density increases. A characteristic case is
shown in  figure~\ref{fig:t5b}.
\end{description}

\begin{figure}[h]
\includegraphics[width=20pc]{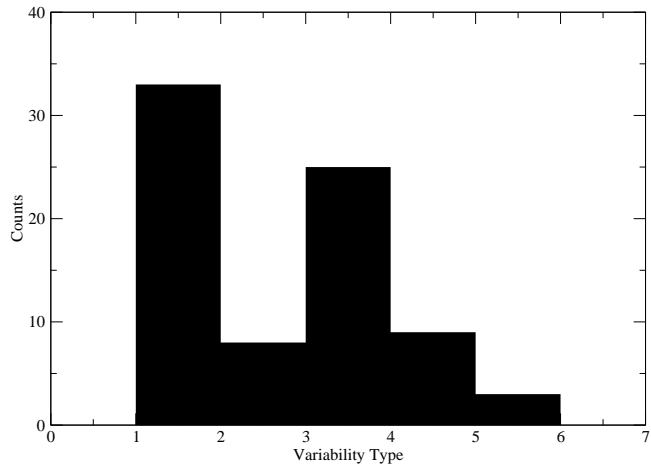}\hspace{2pc}%
\begin{minipage}[b]{14pc}\caption{\label{fig:type_distrb}The distribution of sources over
    spectral variability type. The sub-classes labeled ``b'' are enumerated with appending
  .5 to the class number. For example, 5b should appear as 5.5. The number of sources used
  here are 78.}
\end{minipage}
\end{figure}

The distribution of sources over the different variability types are shown in
figure~\ref{fig:type_distrb}. The sub-classes with the label ``b'' are numbered with
appending the fraction 0.5 to the numerical tag of the corresponding type (e.g. Type 5b
appears as 5.5). As can be seen, the majority of the sources show variability of type 1 --
3b which is the result of the {\em F-GAMMA} sample selection (i.e. sources variable at all
frequencies). Interestingly, a small number of sources (eight) show ``achromatic''
behaviour calling for a dedicated study as it is discussed later.

This classification is done solely on the basis of the phenomenological characteristics of
the variability pattern shown by the radio spectra within a given band-pass. As it is
discussed in the next section, it appears that all the phenomenology for types 1--4b can
be naturally explained by the same physical system observed under different circumstances.

\section{A physical interpretation of the variability types}
In the following it is suggested that the classification discussed earlier can be
reproduced by a simple two-component system observed under different circumstances.

\begin{figure}[h]
\includegraphics[width=17pc]{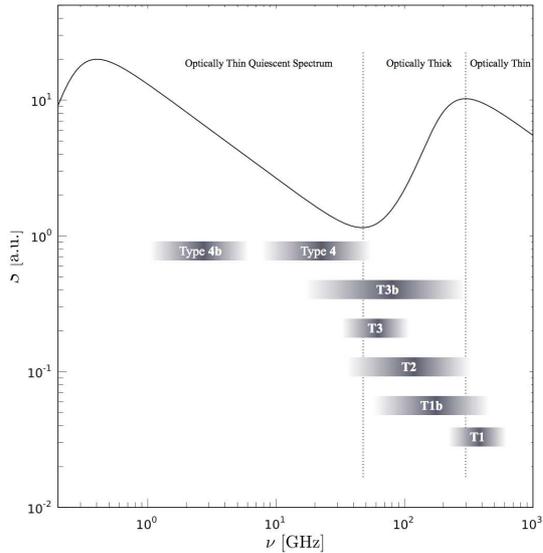}\hspace{2pc}%
\begin{minipage}[b]{14pc}\caption{\label{fig:principal}The assumed common underlying
    two-component system. The different variability types can be reproduced with the
    appropriate modulation of the relative position and relative broadness of the
    band-pass denoted by the grey shaded areas.}
\end{minipage}
\end{figure}

Let us assume a simple two-component system at $z=0$, consisted of:
\begin{enumerate}
\item A power-law quiescent spectrum with $S\propto\nu^\alpha$ (e.g. $\alpha \approx
  -0.5$). This can be attributed to the optically thin diffuse emission of a large scale
  jet or even relic recent flaring event or blends of such.
\item A convex synchrotron self-absorbed spectrum (hereafter SSA) representative of a
  recent outburst superimposed on the quiescent part.
\end{enumerate}
The assumed configuration is presented in figure~\ref{fig:principal} where the shaded
areas denote the observing band-pass. The phenomenology shown there captures the system
(solid line) at an instant in time. Consequently the spectral shape that would be observed
at that instant of time would depend on two parameters of the shaded areas (band-pass)
shown there:
\begin{enumerate}
\item The {\bf \em position} of the shaded areas relative to the high and low frequency
  peak of the assumed system. This parameter denotes the relative position of the centre
  of our band-pass with respect to the source spectrum.
\item The {\bf \em width} of the shaded areas relative to the width of the bridge between
  the optically thick part of the outburst and the steep part of the quiescent
  spectrum. This parameter denotes the fraction of the spectrum that the band-pass can
  sample.
\end{enumerate}
Within this scenario it can be said that as the variability type increases the dominance
of the steep quiescent spectrum becomes progressively larger and the basic characteristics
of the observed variability types 1--4b can be reproduced naturally with the appropriate
modulation of these two parameters.

The question that naturally arises then is how can these two quantities be
modulated. Qualitatively speaking, this can be formulated in terms of the combination of
(a) {\em redshift} and (b) {\em source intrinsic properties}. The {\em redshift} changes
the relative {\em position} of the band-pass allowing a different part of the spectrum to
be sampled. The {\em source intrinsic properties} imply that different sources show
different spectral characteristics, such as: peak frequency of the outburst, peak flux
density excess of the outburst relative to the quiescent spectrum, different broadness of
the valley, different broadness of the SSA spectrum of the outburst etc. Accounting now
for the dynamical evolution of a flaring event in the $S_\mathrm{m}-\nu_\mathrm{m}$ space,
one can introduce a third factor namely (c) the {\em flare specific properties} which of
course are also a function of the {\em source intrinsic properties} and allow the system
to evolve dynamically. While factors (a) and (b) have a static effect and determine the
general shape of the observed spectrum, the latter one (c) changes both the relative
position and width of the band-pass dynamically shaping the specific characteristics of
the variability pattern. In the following we present calculations that have been done to
examine whether this scenario can reproduce the observed phenomenologies.

\section{Modelling the broad band radio spectra}
%
Following the hypothesis that all the observed events are the reflection of the same
process, namely shocks evolving in jets seen with different frequency band-passes at
different evolutionary stages, the Shock-in-Jet model
\cite{Marscher1985ApJ,turler2000AnA...361..850T} has been applied to reproduce their
temporal evolution. 

Specifically, our approach is presented in \cite{2011AnA...531A..95F}, and relies on the
information extracted from the quiescent spectrum and the flaring event to connect the
different radiative evolutionary stages of the shock ({\sl Compton}, {\sl Synchrotron} and
{\sl Adiabatic} stage). The distribution of the quiescent spectrum parameters depends on
the intrinsic source properties such as the magnetic field in the jet, $B$, the Doppler
factor, $D$, and the normalisation coefficient of the spectrum, $K$. The spectral behaviour
of the injected component, depends on the evolution of the physical parameters of the
shock and is parametrized by the exponents in the relations giving $B$, $K$ and $D$ as
functions of the distance along the jet: $ L\propto R^{r}$, $B \propto L^{-r \cdot b}$, $K
\propto L^{-r \cdot k}$, $D \propto L^{-r \cdot d}$, where $r$ is the opening rate of the
jet. In the current work it is assumed that the {\sl Synchrotron} stage is very short
compared to the other two stages (see \cite{2011AnA...531A..95F}).

\begin{figure}[h]
\begin{minipage}{12pc}
\includegraphics[width=12pc]{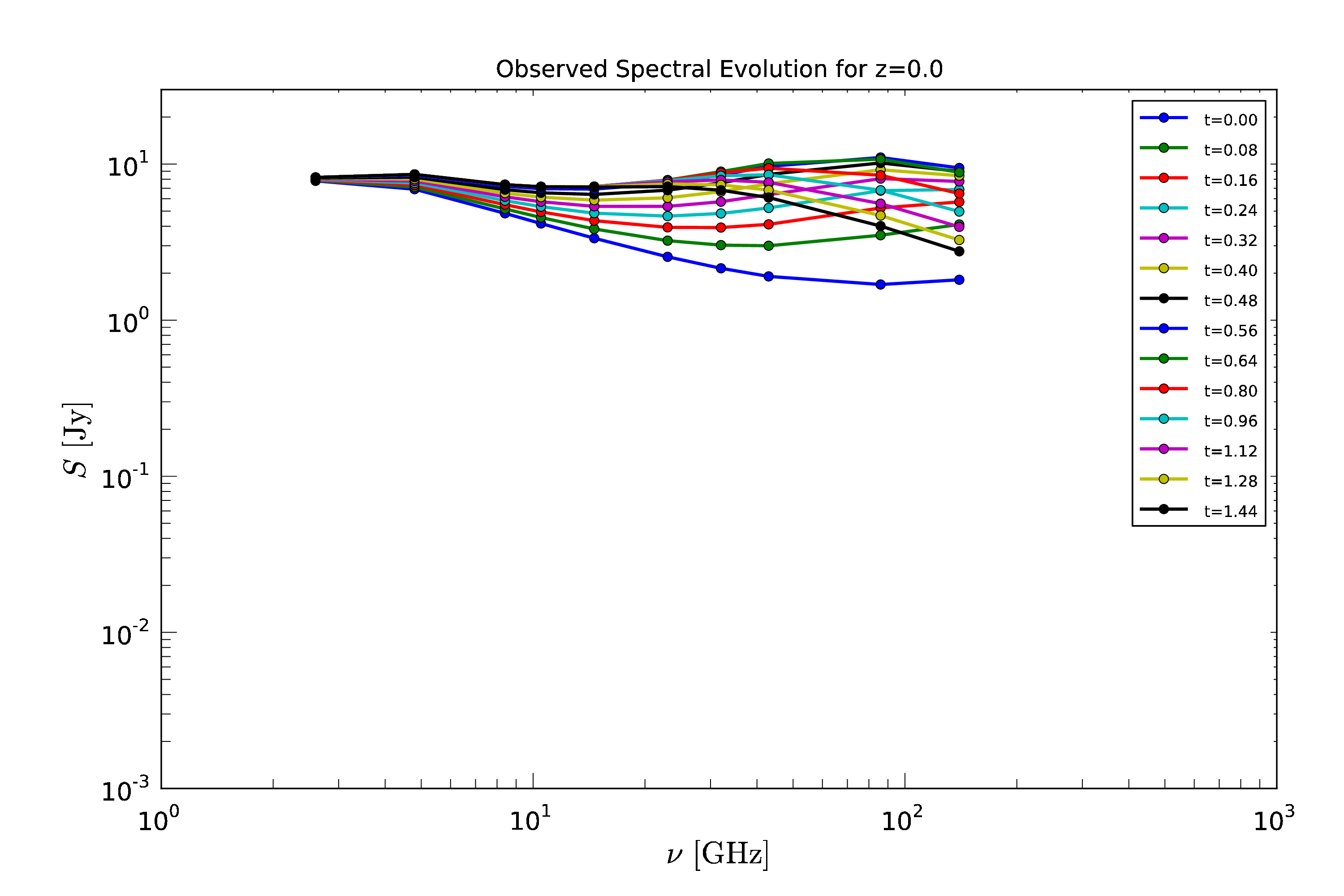}
\caption{\label{fig:high00}The spectrum expected from a powerful source at $z=0$.}
\end{minipage}\hspace{1pc}\vspace{1pc}%
\begin{minipage}{12pc}
  \includegraphics[width=12pc]{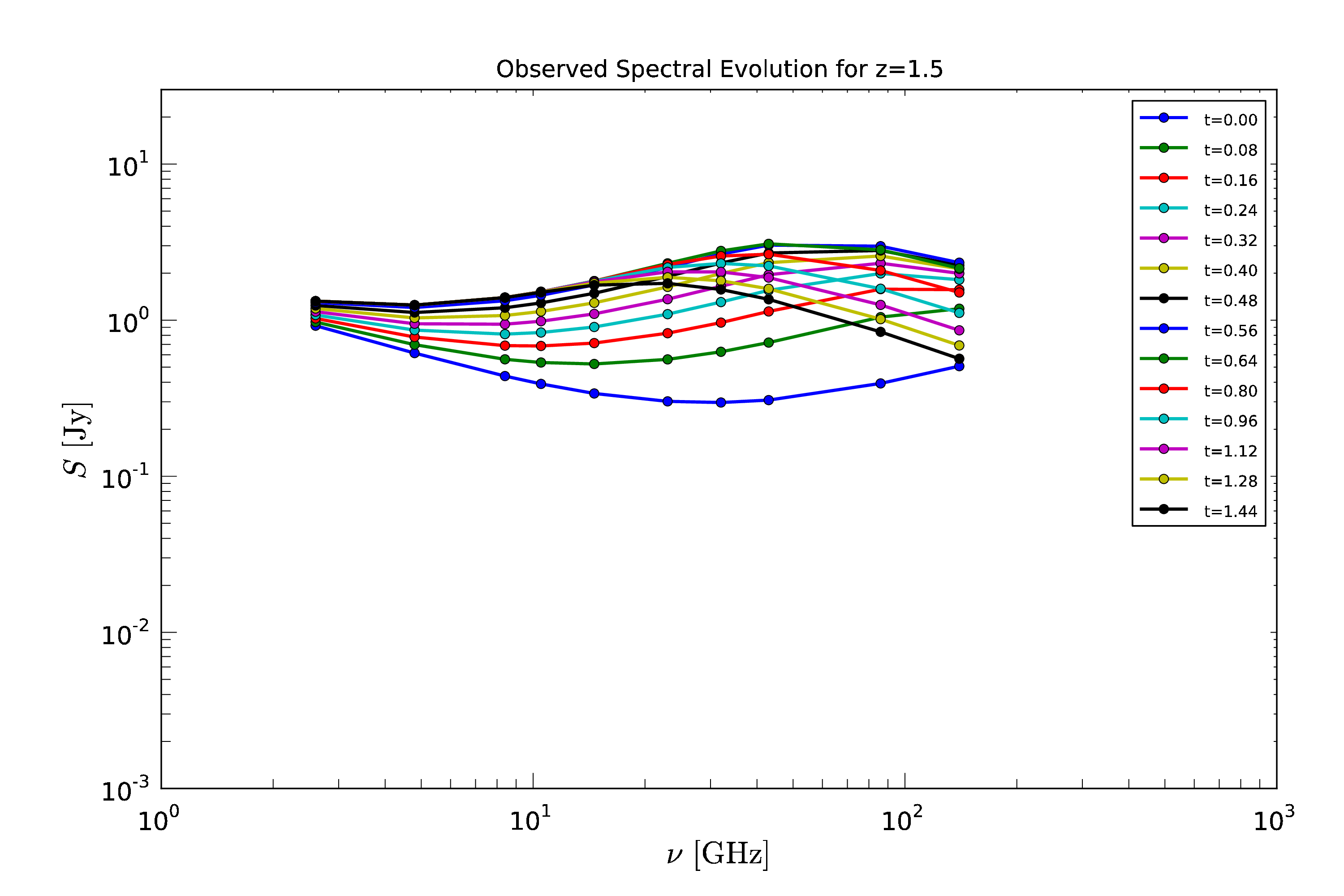}
\caption{\label{fig:high15}The spectrum expected from a powerful source at $z=1.5$.}
\end{minipage}\hspace{1pc}%
\begin{minipage}{12pc}
\includegraphics[width=12pc]{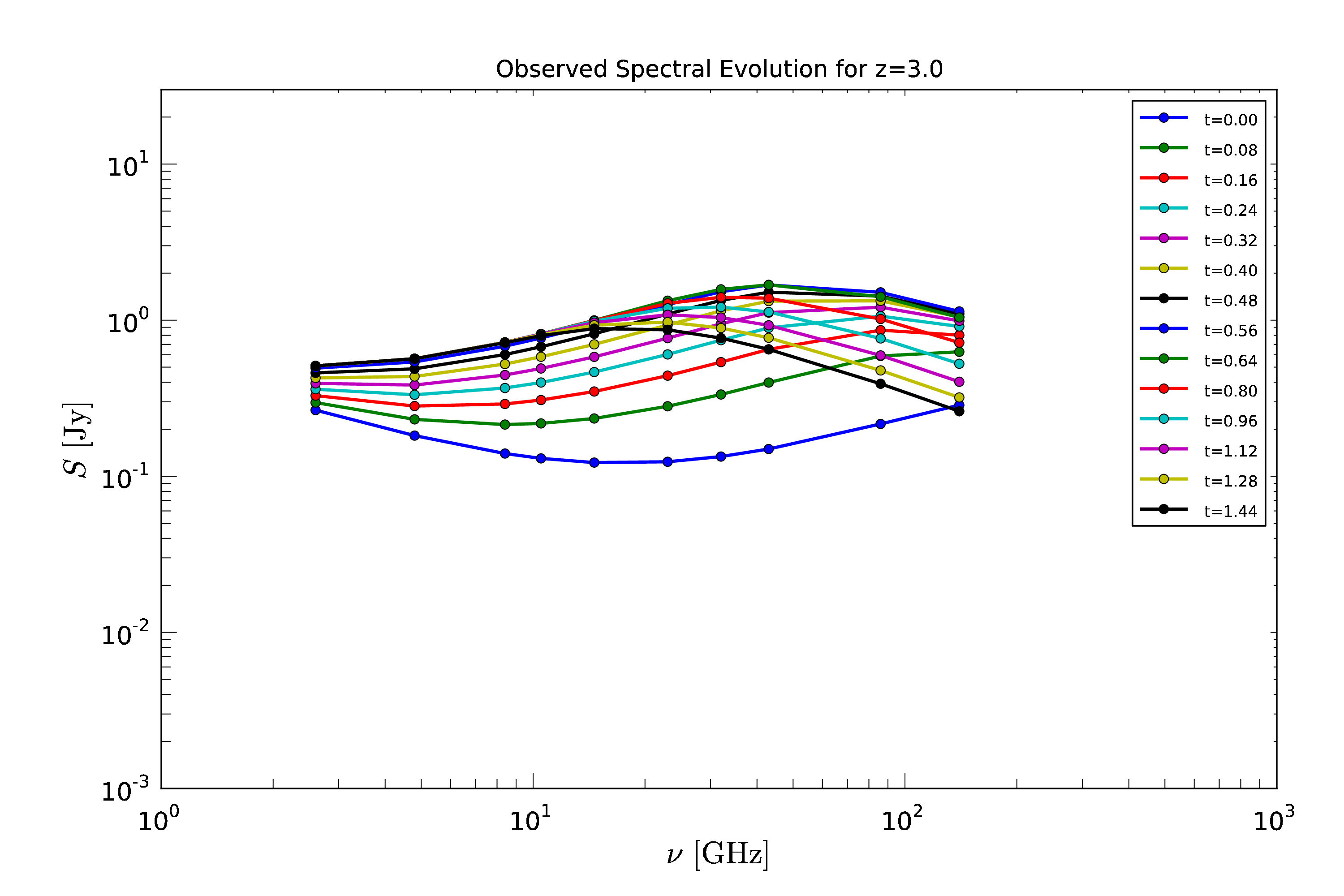}
\caption{\label{fig:high30}The spectrum expected from a powerful source at $z=3.0$.}
\end{minipage}\hspace{1pc}%
\begin{minipage}{12pc}
\includegraphics[width=12pc]{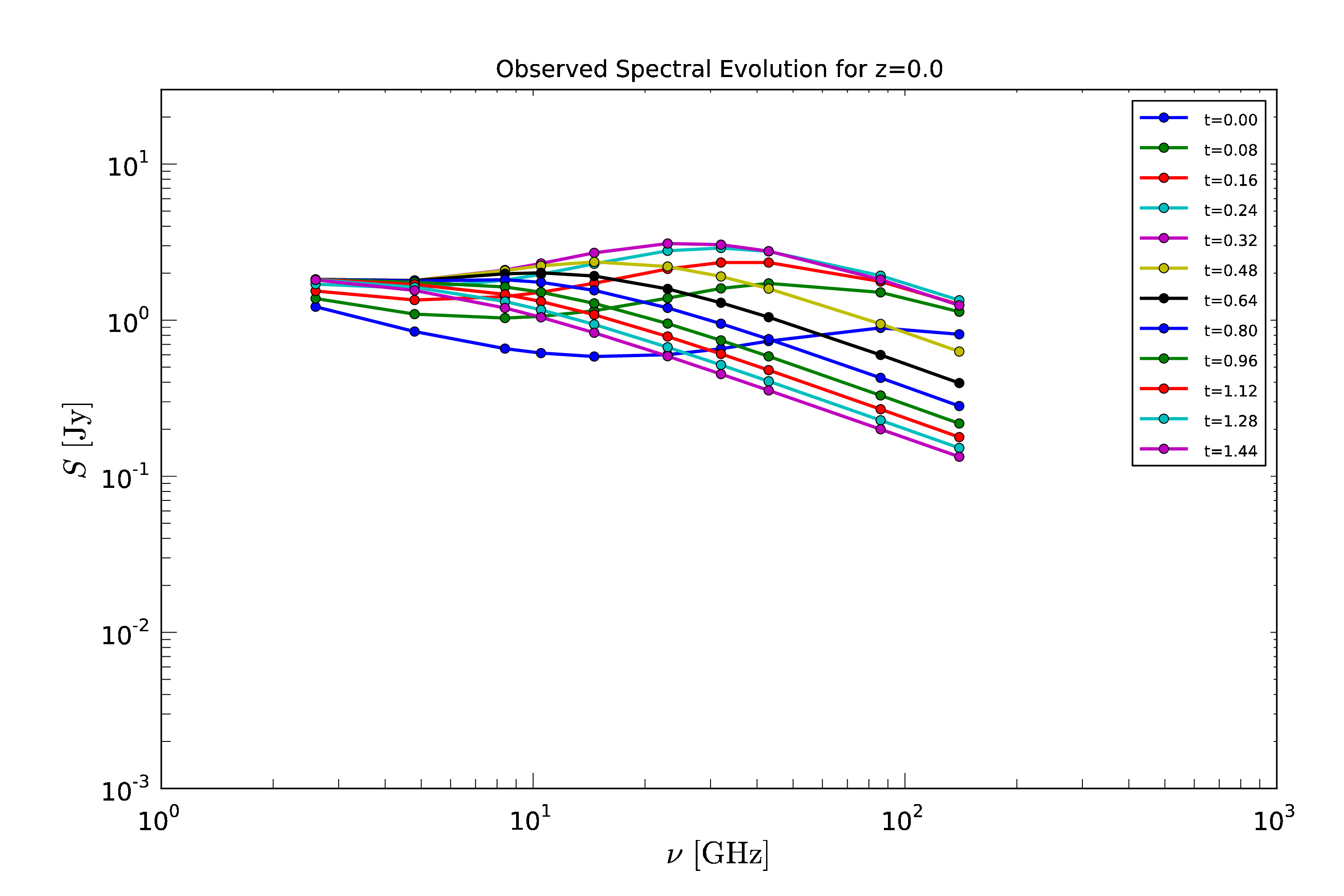}
\caption{\label{fig:med00}The spectrum expected from a medium source at $z=0$.}
\end{minipage}\hspace{1pc}\vspace{1pc}%
\begin{minipage}{12pc}
\includegraphics[width=12pc]{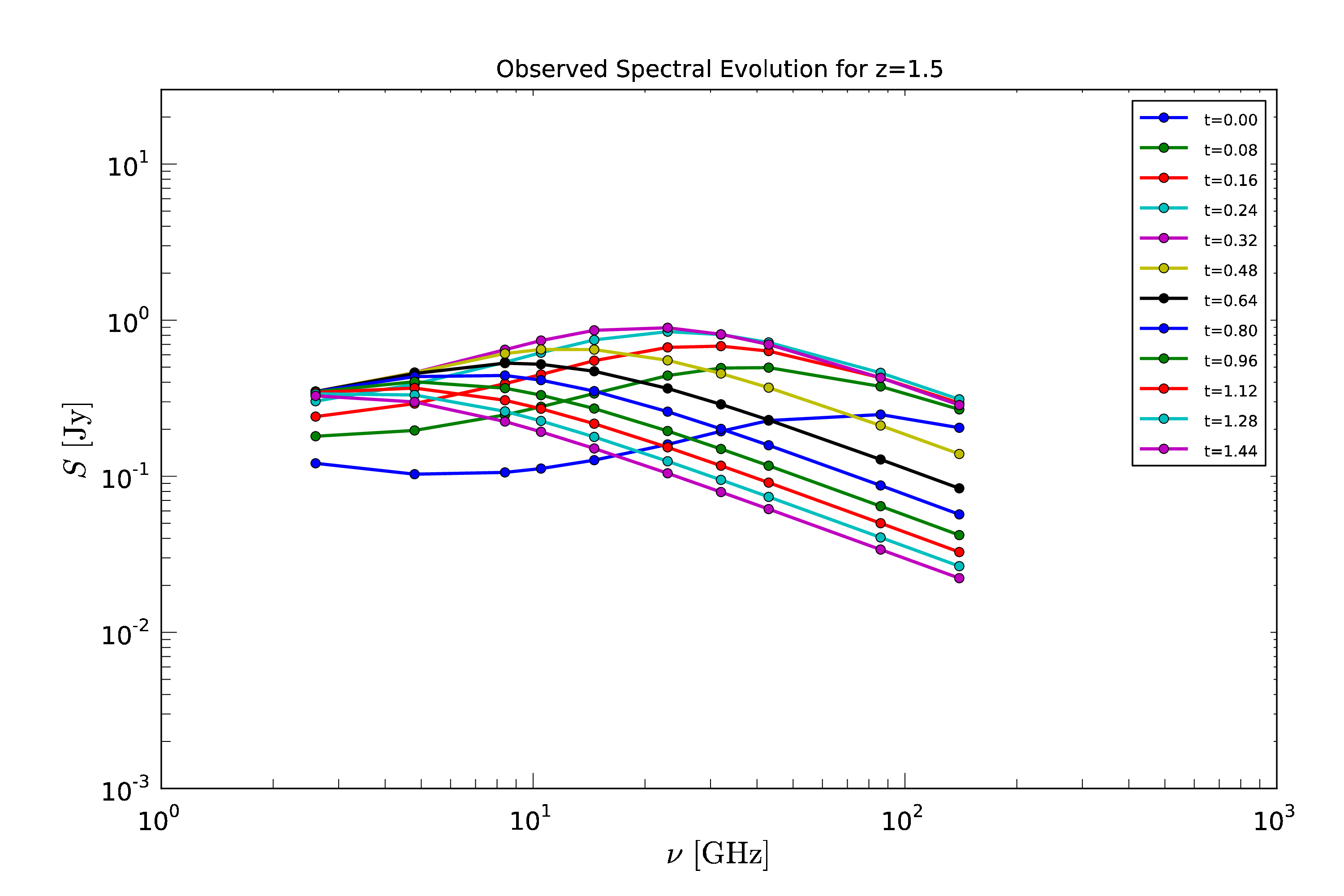}
\caption{\label{fig:med15}The spectrum expected from a medium source at $z=1.5$.}
\end{minipage}\hspace{1pc}%
\begin{minipage}{12pc}
\includegraphics[width=12pc]{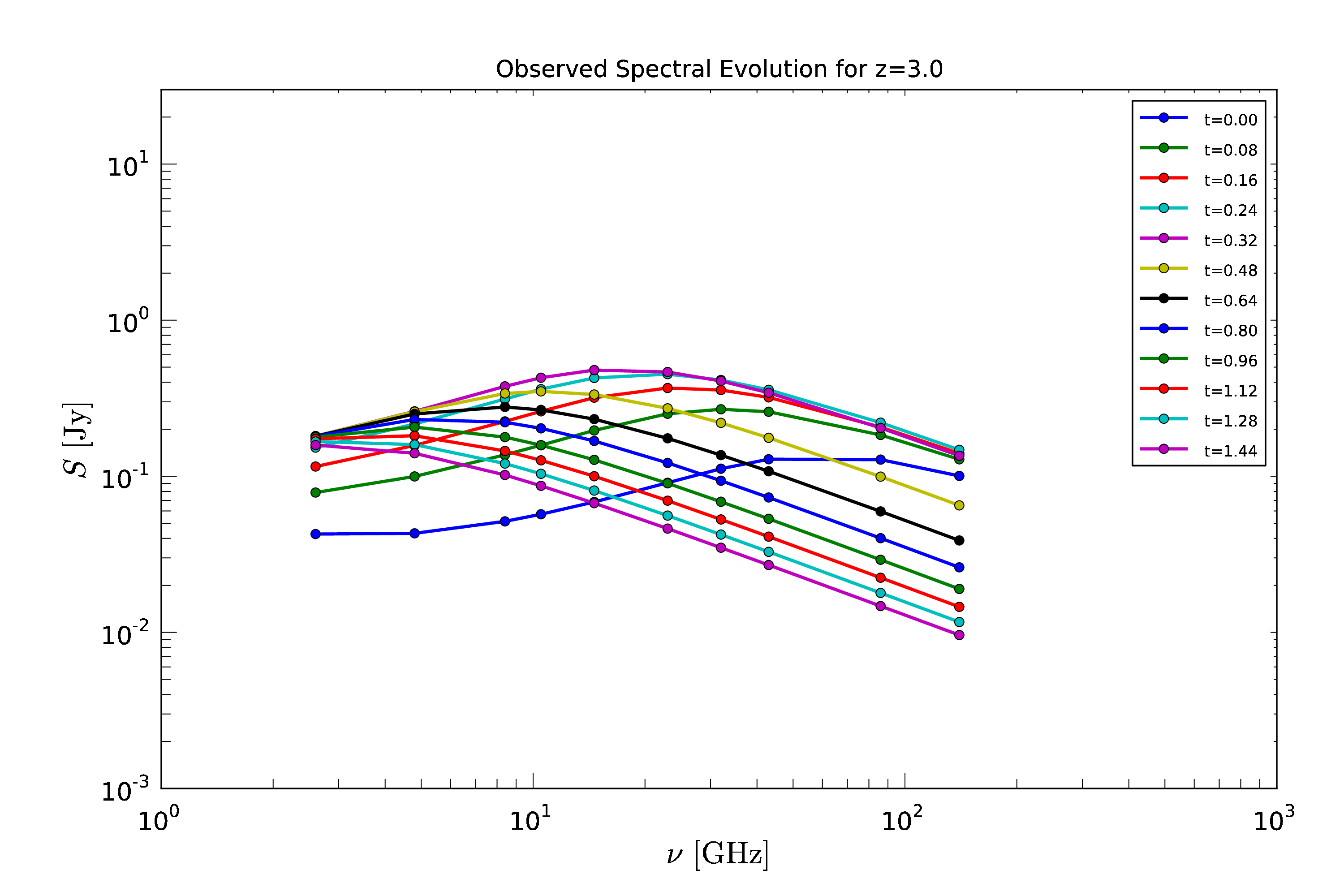}
\caption{\label{fig:med30}The spectrum expected from a medium source at $z=3.0$.}
\end{minipage}\hspace{1pc}%
\begin{minipage}{12pc}
\includegraphics[width=12pc]{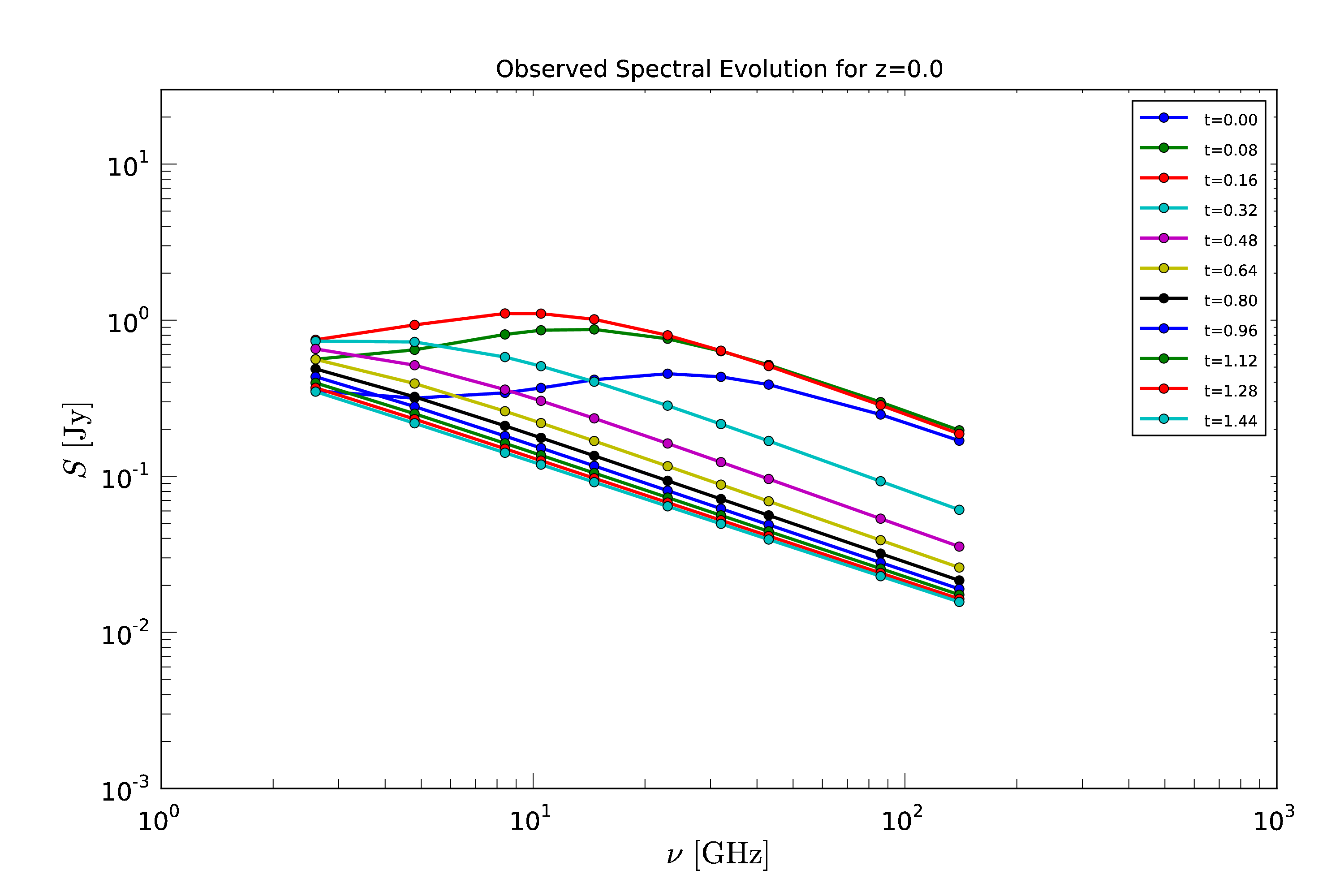}
\caption{\label{fig:low00}The spectrum expected from a weak source at $z=0$.}
\end{minipage}\hspace{1pc}\vspace{1pc}%
\begin{minipage}{12pc}
\includegraphics[width=12pc]{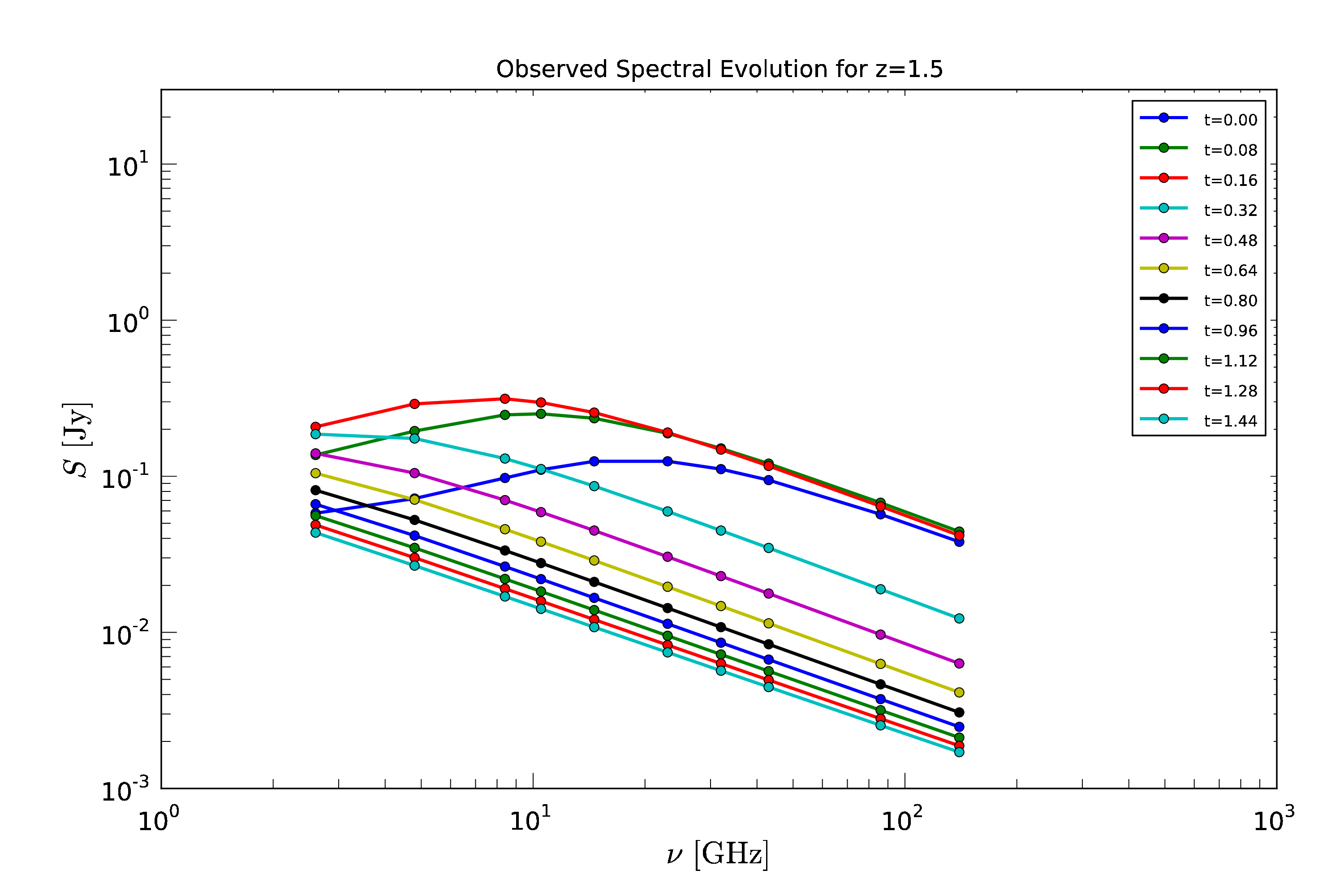}
\caption{\label{fig:low15}The spectrum expected from a weak source at $z=1.5$.}
\end{minipage}\hspace{1pc}%
\begin{minipage}{12pc}
\includegraphics[width=12pc]{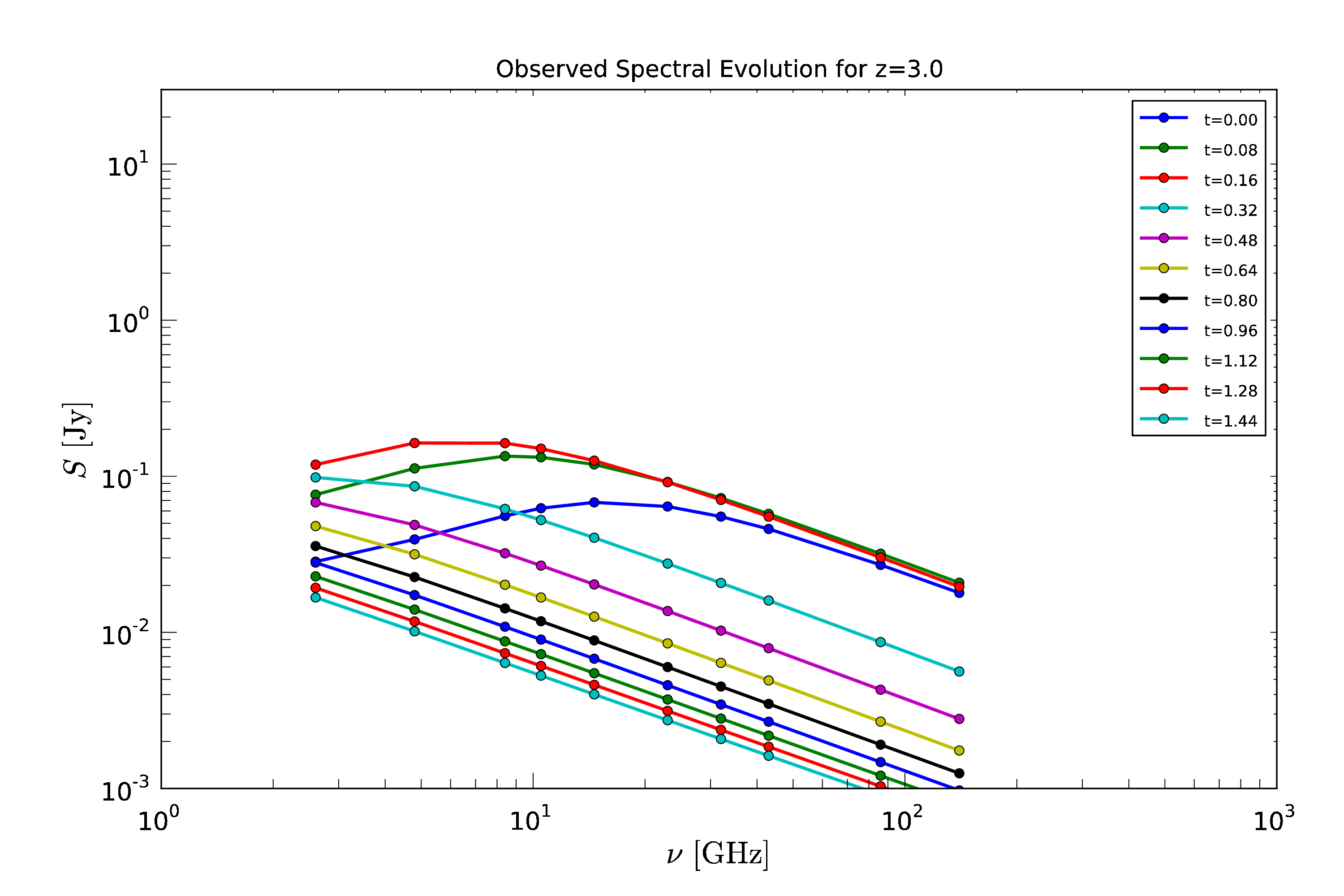}
\caption{\label{fig:low30}The spectrum expected from a weak source at $z=3.0$.}
\end{minipage} 
\end{figure}

We have calculated the spectral evolution of flares for three characteristic cases in
terms of source {\em luminosity}: weak, medium and strong and for three {\em redshifts}:
0, 1.5 and 3. In every case the peak flux of the flaring event at time 0 was set equal to
the turnover flux of the quiescent spectrum. Subsequently the flare has been left to
evolve according to the evolutionary model discussed previously
(\cite{Marscher1985ApJ,turler2000AnA...361..850T,2011AnA...531A..95F}).
Figures~\ref{fig:high00}--\ref{fig:low30} show the spectra expected to be observed by a
facility with band-pass of 2--140\,GHz ({\em F-GAMMA} program). It is evident from there
that the majority of the types discussed earlier are reproduced naturally. Elsewhere
(Angelakis et al. in prep.) more cases are simulated increasing the expected variability
patterns.

\section{Discussion}
From the discussion in section~\ref{subsec:classification}, it is evident that the
variability phenomenologies of the studied blazars can be categorised in two fundamentally
different classes. (a) Sources that are dominated by spectral evolution and (b) sources
that have a convex spectrum and vary self-similarly with only mild if at all spectral
evolution. This implies that there must exist two distinct mechanisms causing
variability. It must be noted though that this refers to the available baseline (roughly
five years) meaning that sources of type 5 and 5b could still show spectral evolution over
longer time scales. An additional element that points towards a different variability
mechanism is the persistency of the spectral shape, in the case of type 5 and the fashion
of change in the case of type 5b which is not seen in cases of clear spectral evolution.

Of the 78 sources that have been examined for the current work, 8 show achromatic
variability. The interesting characteristics is that in the cases that show a mild
spectral evolution, the turnover flux and frequency $S_\mathrm{m}$ and $\nu_\mathrm{m}$,
are evolving in an anti-correlated fashion (see e.g. figure~\ref{fig:t5b}). Apart from the
fact that all of them show the clear presence of a large scale jet even at 2\,cm (as it is
shown from the MOJAVE images \cite{Kellermann2004ApJ}) no other peculiar property has been
identified so far. Possible mechanisms that are currently investigated, include: changes
in the magnetic field structure, changes in the Doppler factors and geometrical effects.

It is noteworthy that none of the studied sources has shown a switch of type at least over
the baseline of the {\em F-GAMMA} program, neither between types of the same underlying
mechanism (i.e. 1--4b) nor between types with different underlying mechanism (i.e. types
1--4b and types 5, 5b). This suggests that the mechanism producing the variability is,
either a fingerprint characteristic of the source, or the conditions that determine it
change over longer time scales, if at all. Possible switch from achromatic to an evolution
dominate behaviour would imply that the suggested dichotomy of the variability mechanisms is
not valid. In any case, the persistency of the evolution dominated types implies that the
power deposited in each event for a certain source is not varying significantly from one
event to the other. Further investigations to explore this statement are underway and will
be presented elsewhere.

Concerning the evolution dominated case, it seems that the Marscher \& Gear model
\cite{Marscher1985ApJ} provides a precise reproduction of the observed phenomenology and
most importantly, over a range of intrinsic parameters covered by the {\em F-GAMMA}
sample. Studies to examine whether other variability mechanisms can reproduce the observed
phenomenology (shapes, time scales etc.) are needed. In any case, any successful model
could be used for extracting the physical parameters from the observed spectra.

\ack Based on observations with the 100\,m telescope of the MPIfR (Max-Planck-Institut
f\"ur Radioastronomie). Based on observations carried out with the IRAM 30m
Telescope. IRAM is supported by INSU/CNRS (France), MPG (Germany) and IGN (Spain).
I. Nestoras and R. Schmidt are members of the International Max Planck Research School
(IMPRS) for Astronomy and Astrophysics at the Universities of Bonn and Cologne. The {\em
  F-GAMMA} team sincerely thanks the Time Allocation Committee of the 100-m and 30-m
telescope for supporting the continuation of the program. E. Angelakis feels obliged to
wholeheartedly thank Dr. A. Kraus for the constant support and the very constructive
discussions.

\section*{References}


\end{document}